\documentclass[a4paper]{article}
\usepackage{graphicx}
\usepackage{listings}
\usepackage{amsmath}
\usepackage{url}
\usepackage{xspace}
\usepackage{a4wide}

\newcommand{\genometool}[0]{{\sf genometools}\xspace}
\newcommand{\chainer}[0]{{\em CHAINER}\xspace}
\newcommand{\coconut}[0]{{CoCoNUT}\xspace}
\newcommand{\Comment}[1]{{}}

\begin{document}

\title{
WinBioinfTools: Bioinformatics  
Tools for Windows High Performance Computing Server 2008
\\
{\large A Joint Project between Microsoft Egypt, Cairo Microsoft Innovation
Center, and Nile University}
}

\author{Mohamed Abouelhoda \thanks{
Center for Informatics Sciences, Nile University, Giza, Egypt
Email: mabouelhoda@nileuniversity.edu.eg, mabouelhoda@yahoo.com
}
\and Hisham Mohamed$^\ast$}

\maketitle

\begin{abstract}

Open source bioinformatics tools running under MS Windows are rare to
find, and those running under Windows HPC cluster are almost
non-existing. This is despite the fact that the Windows is the most
popular operating system used among life scientists. Therefore, we
introduce in this initiative WinBioinfTools, a toolkit containing a
number of bioinformatics tools running under Windows High Performance
Computing Server 2008. It is an open source code package, where users
and developers can share and add to. We currently start with three
programs from the area of sequence analysis: 1) CoCoNUT for pairwise
genome comparison, 2) parallel BLAST for biological database search,
and 3) parallel global pairwise sequence alignment. In this report, we
focus on technical aspects concerning how some components of these
tools were ported from Linux/Unix environment to run under Windows. We
also show the advantages of using the Windows HPC Cluster 2008. We
demonstrate by experiments the performance gain achieved when using a
computer cluster against a single machine. Furthermore, we show the
results of comparing the performance of  WinBioinfTools on the Windows
and Linux Cluster.

\paragraph{\bf Availability:}
WinBioinfTools is open source package available at 
the Nile University Bioinformatics Server
(\url{http://www.nubios.nileu.edu.eg/tools/WinBioinfTools})
and at
CodePlex (\url{http://winbioinftools.codeplex.com}). 
\end{abstract}

\newpage
\section{Introduction}

Recent research in molecular biology requires strong computational
support to cope with the ever increasing amount of biological
data. 
Bioinformatics is the area of science
concerned with the development of
computational tools and methods for the
analysis and management of biological data. 
Currently many biologists
use bioinformatics tools as an essential part of their analysis either to
generate hypotheses before going to lab or to analyze the resulting
data. This practice is clearly 
of great benefit to speed up research
process. However, 
it is in many cases not feasible due to two reasons: First, most
bioinformatics tools are written for Unix/Linux operating systems,
whereas life scientists are familiar more with the windows
environment. Second, these tools are not easily adapted to run under
compute cluster, which hinders the running of compute intensive jobs.

In this paper, we introduce WinBioinfTools, a toolkit containing a
number of bioinformatics software tools running under Windows HPC
cluster 2008.  WinBioinfTools is the result of an interoperability
project between some open source projects (\coconut and others) that
were already developed for the community by Mohamed Abouelhoda
(from Nile University) and between Microsoft Egypt and Cairo Microsoft
Innovation Center. In this pilot phase of WinBioinfTools,  we decided
to start with three basic tools used in analyzing and comparing
genomic sequences. These tools include 

\begin{enumerate}
\item   \coconut, which is used for comparing whole genomic sequences,
\item BLAST, which is used for searching biological databases, and
\item An implementation of the standard sequence alignment algorithm.
\end{enumerate}

In this paper we discuss how the tools mentioned above were ported to
run under Windows and how they were adapted to run over the cluster
architecture. Moreover, we present some experimental results showing
that the performance of the tools on a cluster is superior to that on
a single machine.  Furthermore, we report results about comparing the
performance of the Windows based versions of these tools to the
corresponding Linux versions.

\subsection*{Advantages of using Windows HPC Server 2008}

The experimental results we obtained and our experience in using the
Linux and Windows platform show the following advantages for Windows
HPC server 2008: 

\begin{itemize}
  \item  It is easier to download than Linux based systems; ant out of the box solution with rich and informative documentation
  \item It is very user friendly, even the layman whose focus is on
    using readymade tools and focusing solely on data analysis can
    handle it in a straightforward way. Moreover, the GUI saves user
    time and relieves those users with no computational background
    from writing sophisticated command lines. 
      \item The Windows cluster management tools and the job scheduler
        is very user friendly with intuitive interface and efficiency
        in practical implementation. 
      \item The MS-MPI (Microsoft implementation of MPI) runs smoothly
        through the Visual Studio, i.e., easy to compile and run. It
        also has the feature to run virtually over many
        cores. Furthermore, the debugging features of the Visual
        Studio 2008 which support development of parallel algorithms
        are very helpful and speed up the development cycle, compared
        to the traditional Linux based MPICH2 and the native Linux
        debuggers. 
\end{itemize}

Regarding performance, we noticed that the Windows versions of these
tools are comparatively equal to the Linux versions and in some
instances slightly slower.  The latter observation is in large part
attributed to differences in the underlying IO methods of the
operating system and not to the High Performance Computing
components. (HPC components are basically responsible for
communication between nodes and the job scheduler.)  It is also
attributed to some extent to the fact that the native code of those
applications was written specifically on Linux. Hence, more tweaking
of the code is needed after porting to optimize the performance.

\section{Overview of the tools and algorithms}
\subsection{Sequence alignment}
\begin{figure}[t]
\begin{center}
  \includegraphics[scale=0.7]{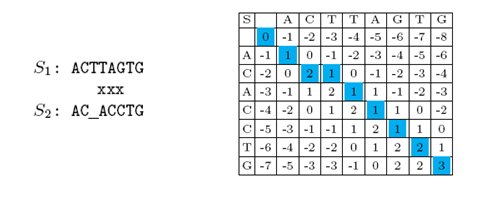}\\
  \caption{Right: Alignment of two sequences. The x symbol indicates
    to a mismatch, and the gaps are represented by the symbol
    "-". Left: The alignment matrix for aligning the two sequences,
    the optimal score is in the lower-left cell, which is 3, and the
    cells on the optimal alignment path are highlighted with blue
    color.} \label{Fig:1} 
  \end{center}
\end{figure}

In computer science terminology, a DNA sequence is a string over an
alphabet of 4 characters, where each character corresponds to a
chemical unit called nucleotide. A protein sequence is also modeled as
a string over an alphabet of 20 characters, where each character
corresponds to a chemical unit called amino acid. A basic problem in
bioinformatics is to find the regions of sequence similarity and
difference among biological sequences. In biology, similar regions
refer to common function and origin, while different regions refer to
traits unique to the respective organism. Biologists call the
differences between sequences mutations, where in a mutation event one
nucleotide (character) replaces another one in the other sequence or
one nucleotide (character) is deleted/inserted from/in a sequence. In
computer science, these types of differences are modeled as edit
operations, where replacements are called mismatches if the characters
are different,  insertion/deletions are called indel (or also
gaps),and common characters are called matches. The computational
model of identifying the differences is called alignment, where one
sequence is drawn on a line and the other on another line such that
the matching/mismatching characters are put above each other, and gaps
are represented by a special symbol, like "-" or "\_". 
Figure \ref{Fig:1} (left) shows an example of two sequences aligned
together. If each of the edit operations is assigned a score, then the
alignment can be scored by summing the individual edit operation
scores. An optimal alignment is the one of highest score. For example,
if a match scores 1, a mismatch scores -1, and a gap scores -1, then
an optimal alignment will maximize the characters matching each
other. 

The algorithm of Needleman and Wunsch \cite{NEE:WUN:1970} 
is based on the dynamic
programming paradigm to efficiently compute an optimal alignment for
two sequences, where an optimal alignment can be decomposed into
optimal sub-alignments. More precisely, for two sequences $S$ 
(of length $n$) and $R$ 
(of length $m$), 
the alignment at position $i$ in $S$ and
$j$ in R can be computed in terms of optimal alignment at
positions $i-1$ and $j-1$.  
Let $\alpha$ denote the cost of
a match, $\beta$ denote the cost of a mismatch, and $\delta$ denote
the cost of a gap, and let $A(i,j)$ denote the score of an
optimal alignment ending at position $i$ in $S$ and $j$ in $R$.
Let $A(i,0)=-\delta i$, $A(0,j)= -\delta j$, for all
$0 \le i \le n$, $0 \le i \le m$, the following recurrence computes an
optimal alignment.

\begin{displaymath}
A(i,j) = \max \left\{ \begin{array}{ll}
  A(i-1,j-1)+\alpha & \textrm{if $S[i]=S[j]$}\\
  A(i-1,j-1)+\beta  & \textrm{if $S[i] \ne S[j]$}\\
  A(i-1,j)+\delta   & \\
  A(i,j-1)+\delta & 
  \end{array} \right.
\end{displaymath}

The intermediate optimal scores of this recurrence can be stored in a
table (2D alignment matrix), where an entry $(i,j)$
tabulates the optimal score at positions $i$ in $S$ and $j$ in $R$. 
The matrix is filled starting from the cell $(0,0)$ in a row-wise,
column-wise, or diagonal-wise fashion until reaching position
$(n,m)$, where the optimal alignment score is
retrieved. 
Note that according to the recurrence given above,
computing the score at position $(i,j)$ in the matrix
depends on the score at positions $(i,j-1)$,
$(i-1,j)$, $(i-1,j-1)$. 
Figure \ref{Fig:1} (right) shows an example of an alignment matrix.

The running time and space consumption of this algorithm is quadratic,
and a computing cluster provides a good solution to this problem,
especially when memory consumption presents a bottle neck. 

\begin{figure}[!t]
\begin{center}
  \includegraphics[scale=0.7]{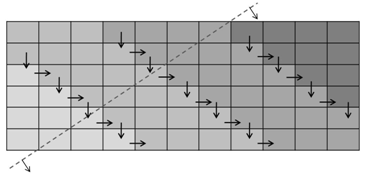}\\
  \caption{The alignment matrix, where the cells of the same color are
    assigned to one of the four cluster node. The dashed diagonal line
    corresponds to the values to be filled in one iteration.  The
    black arrows correspond to values that must be transferred from
    one node to another in order to complete the computation, since in
    the recurrence each cell requires the values in additional three
    cells $(i-1,j)$, $(i,j-1)$, and $(i-1,j-1)$. 
    \label{Fig:2}} 
  \end{center}
\end{figure}

The idea of a cluster based algorithm is to divide the alignment
matrix over the cluster nodes and let each node compute the recurrence
for that portion of the matrix \cite{GAP:BRU:1990}. 
The matrix can be filled in a
diagonal-wise fashion, where the master node can synchronize the
diagonal moves between all nodes. Note that at the boundaries of the
matrix portions, some nodes need to exchange messages to find out the
values in certain cells. Figure \ref{Fig:2} 
illustrates the process of filling
the alignment matrix over four cluster nodes.

\subsubsection*{Parallel sequence alignment on Windows HPC Server 2008}
We developed a version of the global sequence alignment algorithm that
runs under cluster platform. Note that this type of using cluster
platform is categorized as compute cluster where the cluster nodes
cooperate and exchange messages to solve the problem. We used MPI to
share values among the cluster nodes and to synchronize the tabulation
process. We implemented a Linux based version and a Windows based
version and compared the performance of one version to the other on a
single machine and on a cluster of four nodes, see 
Section \ref{SEC:ExperimentalResults}.

\subsection{\coconut}
The genome is the total DNA in a cell. In higher organisms, like
humans, the genome is divided into many units called
chromosomes. Genome comparison boils down to identifying regions of
similarity and difference among two or multiple genomes, where the
common regions should have common function and the different regions
can refer to features unique to the respective genome. The task of
comparing genomes is complicated by three factors: 

\begin{itemize}
  \item Large genome sizes, where the simplest bacterial genomes are
    in the range of Mega characters and higher organisms can reach the
    multi Giga character range. 
  \item Rearrangement events, where segments of the genome change
    order and direction. 
      \item Mutation events, where the similar regions might contain
        character replacement, and/or deletion/insertion. 
\end{itemize}

To cope with the large genomic sequences, \coconut
\cite{ABO:KUR:OHL:2008} 
finds regions of
high similarity using the anchor-based strategy that is composed of
three phases: 
\begin{enumerate}
  \item    
    Computation of fragments (short exact matches among genomic
    sequences). This step is efficiently carried out using the
    enhanced suffix array (implemented in third party packages like
    Vmatch or \genometool). 
    \item Computation of highest-scoring chains of collinear fragments
      (chains are defined below). Each of these highest scoring chains
      corresponds to a region of similarity. The fragments in each of
      such chains are the anchors. These chains are efficiently
      computed using the program \chainer \cite{ABO:OHL:2004}. 
    \item Post processing the resulting chains, which includes, e.g.,
      visualizing and aligning the regions between the anchors of a
      chain using the standard dynamic programming algorithm. 
\end{enumerate}

The rationale of generating fragments is that the regions harboring
fragments are regions of potential similarity, while those regions
containing no fragments are different and are excluded from further
processing. To assert that the fragments actually constitute regions
of similarity and did not appear by chance, they are further processed
by the chaining algorithms. 

The chaining step is the core of the \coconut system. Here we explain
more about the chaining problem and the algorithm solving it. Given
two genomes, a fragment is a similar region between the two
genomes. More precisely, we use fragments of the type maximal exact
matches defined as follows. If $S_1[1..n]$ is the string representing the
first genomic sequence and $S_2[1..n]$, a maximal exact match is a pair
of substrings $S_1[i_1..j_1]$ and $S_2[i_2..j_2]$
such that the characters of these two strings are identical 
(i.e., $S_1[i_1..j_1]=S_2[i_2..j_2])$ and
$S_1[i_1-1] \ne S_2[i_2-1]$ and $S_1[j_1+1] \ne S_2[j_2+1]$, i.e.,
the match cannot be extended to the left and to the right in the two
strings simultaneously. For a fragment $f$  corresponding to the maximal
exact match of the substrings $S_1[i_1..j_1]$ and
$S_2[i_2..j_2]$, we write 
$beg(f)=(beg(f).x, beg(f).y)=(i_1,i_2)$ 
to denote the position at which the fragment
starts in the first genomic sequence and write 
$end(f)= (end(f).x, end(f).y) =(j_1,j_2)$ 
to denote the position at which the
fragment ends in the second genomic sequence. In a two dimensional
space, where the first sequence is the $x$-axis and the second sequence
is the $y$-axis, a fragment can be represented by a rectangle whose
extreme corner points are $beg(f)$ and $end(f)$. 
In the
first phase of \coconut we compute all maximal exact matches of length
at least $l$ using the \genometool package. 
 
A chain of fragments is a sequence of fragments
$fi_1,fi_2,\dots, fi_t$  
such that for any two successive fragments $f^{'}$ and
$f^{"}$, we have $end(f^{'}).x < beg(f^{"}).x$
and
$end(f^{'}.y < beg(f^{"}).y$. 
The score of this chain is
$score(C)= \sum_{k=1}^{n}w(fi_k)$, 
where $w(f)$
is its length, i.e., $w(f)=end(f).x-end(f).x$. The chaining
problem (also known as sparse dynamic programming problem) is to find,
given a set of fragments, a chain C with highest score. The program
\chainer can find this highest scoring chain using techniques from
computational geometry. It uses the line sweep paradigm, where the
fragments are scanned with respect to one dimension. When a start
point of a fragment is scanned we search for a fragment at which a
highest scoring chain ends. The end point of this fragment should lie
in the region bounded by the origin point $(0,0)$ and the currently
scanned start point. This search is carried out efficiently by
arranging the end points of the fragments in a kdtree and using a
range maximum query over it, see \cite{ABO:OHL:2004,ABO:KUR:OHL:2008} 
for more details. 

\subsubsection*{Porting \coconut to run on Windows HPC Server 2008}
\coconut was originally written to run on Linux. Therefore, we need to
port it to the Windows environment. The porting of \coconut involves
three modules: 1) fragment generation, 2) chaining, and 3) and plot
generation to the windows environment. These modules are integrated in
a single program interface written in Perl. The Perl based interface
handles options and module parameters, and performs also input/output
format transformation needed to integrate the different modules. The
details of porting the three modules to Windows are as follows. 
\begin{enumerate}
  \item  The first module is based on a third party package called
    \genometool, which is open source package available at
    \url{www.genometools.org}.  For compiling and building this package to
    run under windows, we investigated the application of the SUA tool
    distributed with the Windows HPC Server 2008 and open source
    package Cygwin.   Using Cygwin, the \genometool package could be
    compiled with minor modifications in the compilation options
    (basically we switched off the fpic option). Once compilation is
    completed, \genometool can run through either the Cygwin or SUA
    interface. (SUA can be used to compile Linux/Unix native
    applications, but this did not work smoothly with \genometool,
    because of some missing libraries.) 
    \item The second module is based on the program \chainer, which was
      readily compiled using SUA and Cygwin, or any gcc compiler under
      Windows. A version of it was also compiled using Visual Studio. 
    \item The third module currently visualizes the resulting chains
      by producing a 2D plot of the comparison. This visualization
      step is achieved by running the open source gnuplot program. In
      next version, we plan to add a module for producing detailed
      alignment. 
\end{enumerate}

The correctness of the porting steps was verified by comparing the
results of the windows version to the Linux based version. Performance
comparison over a single node and a cluster of four nodes was also
done and the results will be reported in the next section. 

The system \coconut was wrapped to run under Windows cluster to
distribute all pairwise comparisons, given two multi-chromosomal
genomes, on the cluster nodes (i.e., running in a load balancing
mode). The job scheduler of the Windows HPC Server 2008 was used as
the underlying cluster engine to achieve this task.  Another version
of the cluster based \coconut was implemented to run under Linux
Cluster. We compared the performance of the two versions of \coconut by
comparing human chromosomes to mouse chromosomes. The two versions run
on the same hardware. The result of this comparison is given in the
results section.

\section{BLAST}
A biological database is a collection of DNA or protein
sequences. Given a biological sequence, usually a short one, the
program BLAST is used to find sequences similar to it. Note that
searching a database can be done using the standard alignment
algorithm mentioned above, where the query sequence is aligned to each
sequence in the database. But the running time is prohibitive, so the
program BLAST \cite{ALT:GIS:MIL:MYE:LIP:1990}, 
which implements a heuristic algorithm, is used for
this task. BLAST finds the similar sequences in a time proportional to
the database length and it is very fast in practice. BLAST is the most
popular software tool in Bioinformatics and it is available as open
source.

In biological applications, the program BLAST is required to answer
millions of queries, emerging either from remote users or from certain
newly sequenced DNA/protein segments. Since the time in which BLAST
finds similar sequences to a query is proportional to the database
length, it is ideal to use compute clusters to divide (segment) the
database and distribute each query job over the cluster nodes; the
results are then collected back to the master node \cite{DAR:CAR:FEN:2003}. 

\subsubsection*{Parallel BLAST on Windows HPC Server 2008}

We developed a wrapper for BLAST (called parallel BLAST) that runs
under Windows cluster. Specifically, we wrote a program that performs
database segmentation and distributes the queries over the cluster
nodes. We implemented two versions of this program to run under the
Windows cluster: one using MPI and one using the job scheduler. We
also implemented another version running under Linux and based on the
MPI. In fact, handling BLAST queries can be done through the job
scheduler with no need to write a specific MPI program. But the reason
why we used MPI here is to measure communication time for answering
queries, a parameter that cannot be measured through job
schedulers. The result of comparing the two versions based on MPI is
given in the section about the experimental results. 

\section{Experimental results}
\label{SEC:ExperimentalResults}
We performed experiments to compare the running time of the Windows
cluster versions of our programs to the corresponding versions running
under Linux Cluster. The experiments we performed addressed the load
balancing feature of the Windows cluster by testing the Windows job
scheduler and they also addressed the compute cluster feature by
testing the Microsoft MPI implementation against the open source MPI
implementation MPICH2. We used Visual Studio 2008 for compiling MPI
based applications with the option "Profile Guided Optimization" set
on. The programs were executed on Windows with the option "-affinity"
set on in the job submission command line. The hardware setting for
the experiments is given in Appendix I. The cluster setting for
Windows is as follows: Windows HPC Server 2008 (production version),
MS-MPI with Network direct for running MPI applications, affinity was
set to 8 on each node (we have 2 Quad core processors).  (We did not
use OpenMPI in our code.)  The Linux cluster runs on the same
hardware, where the Windows cluster runs (dual boot configuration). We
used openSUSE Linux 10.3 as Linux operating system and used MPICH2
v.1.0.8 for MPI. We used MPD as process manager within MPICH2.  In the
coming subsection, we present the results of testing the load
balancing feature. Then in the next subsection we show the results of
testing the MPI feature.

\begin{table}[!t]
\begin{center}
\begin{small}
\begin{tabular}{|c|c|c|c|}
\hline
{Database} & {Type} & {Description}  & {Size (GB)} \\
\hline
\hline
Drosph    & DNA    &   Sequences from the drosophila genome    & 0.12 \\ \hline
pataa     & Protein   & Patented protein database               & 0.17 \\ \hline
est\_others & DNA      &  EST database without human/mouse entries &0.37 \\ \hline
env\_nr     &  Protein &Environmental protein sequences &          1.6 \\ \hline
nr          & protein  &Non-redundant protein database    &       4 \\ \hline
\end{tabular}
\end{small}
\end{center}
\caption{\small Table 1: The Biological databases used to test the program BLAST} \label{Tab:1}
\end{table}

\subsection{Testing load balancing features of the Windows HPC Server 2008}

    \subsubsection*{Querying BLAST} 
    Table \ref{Tab:1} shows the different databases
    we used to test the parallel version of BLAST. From each database,
    we sampled 1000 query sequences; these were taken at random from
    the database. We then launched each query back over the respective
    database. The running times are the time of answering the total
    1000 query. This experiment was conducted over a single node and a
    cluster of four nodes, see Table \ref{Tab:2}. The communication time was
    the time needed to send the queries to the four cluster
    nodes. From the table, we can see that the running times on both
    operating systems are comparable, and the Linux version is faster
    for larger databases. But we think that this advantage of the
    Linux Cluster becomes less relevant as more nodes are added to the
    cluster due to the following. The query time is proportional to
    the database size and the higher the number of nodes, the less
    database size on individual nodes, and the more reduced query
    time, which will ultimately lead to almost equal running
    time. Unfortunately we could not verify this claim, because the
    number of nodes available at CMIC is four.

\begin{table}[!t]
\begin{center}
\begin{small}
\begin{tabular}{|c||c|c||c|c||c|c||c|c|}
\hline
Database &      
\multicolumn{2}{c||}{One node}   
& \multicolumn{2}{|p{2cm}||}{4 nodes with com. time}
& \multicolumn{2}{|p{2cm}||}{4 nodes without com. time}
& \multicolumn{2}{|c|}{com. time} \\ \hline 
& Win &  Linux & Win &  Linux 
& Win &  Linux & Win &  Linux \\ \hline \hline
Drosoph &   
0.08 &  0.06 & 0.038 & 0.023 & 0.0235 & 0.0188 & 0.0145 &
0.0042 \\ \hline
Pataa  &    
0.5  & 0.4 &  0.1344 &  0.1   &  0.0184 & 0.014 & 0.116  & 0.086
\\ \hline 
est\_others &
1    &  0.8 &  0.5799 &  0.421 &  0.0343 &  0.035 & 0.5456 &  0.386
\\ \hline
env\_nr &
18   &  15  &  4.0308 & 3.5132 &  0.5308 & 0.5132 & 3.5    &   3
\\ \hline 
Nr  &   27  &  24      & 7.2077 &  6.1163 & 0.4077 & 0.6163 &   6.8 &  5.5
\\ \hline 
\end{tabular}
\end{small}
\end{center}
\caption{
\small
The running time (in hours) of 1000 query on one and four
nodes. 
The clolumn titled ``One node'' contains time on a single machine.
The column titled `` 4 nodes with com. time'' reports running times on
four cluster nodes including
communication time. 
The column titled ``4 nodes without com. time'' reports running times on four
cluster nodes 
excluding communication time. 
The column titled ``com. time'' reports communication times only on
four cluster nodes. 
The columns titled ``Win'' and  ``Linux'' contains running times for Windows
and Linux, respectively.
\label{Tab:2}
}
\end{table}

    \subsubsection*{All against all comparisons using \coconut} 
\begin{figure}[!t]
\begin{center}
  \includegraphics[scale=0.61]{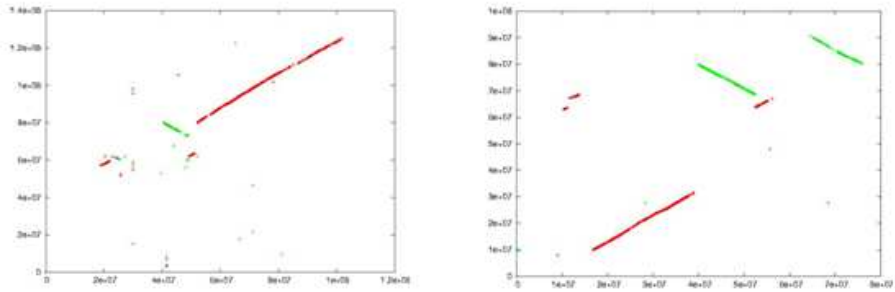}\\
  \caption{
    \small
    Left: Comparison of the human Chromosome 13 to the mouse
    Chromosome 14. Right: Comparison between the human chromosome 18
    and the mouse chromosome 18. Red lines show similarity in the
    forward direction of both chromosomes, while green lines
    correspond to similarity between the forward strand and the
    backward strand.} \label{Fig:3} 
  \end{center}
\end{figure}

    We
    compared some of the human chromosomes to some of the mouse
    chromosomes. We have a total of 20 comparisons. Figure \ref{Fig:3}
    contains the resulting 2D plots showing the common regions
    between some chromosomes; these plots were automatically
    generated based on gnuplot. On the windows machine, the total
    time on a single node was about 47 hours, while on four nodes it
    was about 12 hours.  On the Linux machine, the total time on a
      single node was about 40 hours, while on four nodes it was 9
      hours. The jobs were submitted based on a shell script to run
      the program on each node over the designated data, which were
      pre-distributed over the nodes. This time difference between
      Linux and Windows versions is basically attributed to the
      fragment generation phase done by \genometool, which must be
      compiled under Cygwin. Although we ran the resulting code using
      the SUA component of the Windows server 2008, the running time
      was still higher than that of the Linux version. We think that
      this is attributed to some wrapping done by Cygwin/SUA around
      \genometool functions. In coming Windows version of \coconut, we
      plan to replace \genometool with our own implementation that
      compiles and runs directly on Windows without any intermediate
      layer to reduce the time difference.

\subsection{Testing the MPI implementation of the Windows HPC Server 2008}


The task of sequence alignment, as mentioned before, requires that the
cluster nodes participate in tabulating the optimal partial results
until reaching the final optimal result. We implemented a parallel
version of this algorithm using MPI. We generated random sequences of
different sizes and compared the Windows cluster based implementation
against the Linux based implementation. Table \ref{Tab:3} shows the running
times in seconds for different sequence lengths and with/without the
communication time. From the table, it can be seen that the version on
the Linux cluster is slightly faster, and this is not attributed to
the underlying cluster technology but to the time on the single
nodes.

\section{Conclusions and future work}
Windows HPC Server 2008 is a very useful tool for research and
development labs. It gives   researchers the opportunity to actually
focus their time and effort on research itself, leaving all the
technical details and setup for the system. Its ease of download and
installation underlines the "out-of-the-box reputation" it has, unlike
Linux systems, where command line and script knowledge is needed to
setup the system and the applications running on it. Also, running a
job on Windows HPC server 2008 can be achieved through an intuitive
GUI, which leaves time for the researchers to focus on the analysis.

For developers building parallel programs, the debugging capabilities
of Visual Studio 2008 are very attractive. They allow and support the
development of parallel algorithms and speed up the development cycle
of new applications.

As for the speed of execution on Windows HPC server 2008, it is
comparative to that on Linux. In our specific experiments, the results
show that the speed on Windows HPC server is impacted, in some
instances, by the use of intermediate layers instead of having a code
that compiles directly in Windows and by intensive I/O and memory
access.

\subsection*{Extension and follow-up work}

After completing this stage of the project, we think that the
following extensions will improve the performance and usage of the
tool. 
\begin{itemize}
\item It is planned to build a standalone fragment generation package
  with multi-core and cluster architecture support to replace or
  enhance the currently used \genometool. This package will be faster,
  light weight, and run directly on Windows without the need to any
  middleware, like SUA or Cygwin. This tool can be provided to the
  users as a standalone package, which could be used in applications
  other than \coconut. 
\item It is desirable to provide other options of \coconut such as
  detailed alignment on the character level, multiple genome
  comparison, repeat analysis, and EST/cDNA mapping.. We hope to
  develop these features in a future version. 

\end{itemize}

\begin{table}[!t]
\begin{center}
\begin{small}
\begin{tabular}{|c||c|c||c|c||c|c||c|c|}
\hline
Length &      \multicolumn{2}{c||}{One node}   
& \multicolumn{2}{|p{2cm}||}{4 nodes with com. time}
& \multicolumn{2}{|c||}{com. time} 
& \multicolumn{2}{|p{2cm}|}{4 nodes without com. time}
\\ \hline 
& Win &  Linux & Win &  Linux 
& Win &  Linux & Win &  Linux \\ \hline \hline
  $100^2$ &   0.0034 &  0.0023 & 0.0018  &   0.001 &  0.0362 &
  0.0161  & 0.0007     &     0.0001 \\ \hline
 $1000^2$  &  0.04  &  0.039 &  0.014 &  0.01 & 0.1527 &  0.0177 &
 0.005 &   0.003 
 \\ \hline
 $5000^2$  &   3.9  &  2.67  &     1  &  0.8  & 0.1423 &  0.152  &
 0.3   &   0.2
 \\ \hline
$10000^2$ &   8.4  &    7   &   2.6  &  2    &  1.19  &  1.28   &
1.1   &   0.7
 \\ \hline
$20000^2$ &   18   &    16  &    8   &  7    & 3.679  &  3.7939 &
2     &   2
 \\ \hline
$30000^2$ &   40   &    37  &     15 &  13.3 &     4  &  3.8    &
11    &   9.5
\\ \hline 
\end{tabular}
\end{small}
\end{center}
\caption{
\small
The running times (in seconds) for pairwise
sequence alignment on one and 4 nodes. In the first column, we
list the sequence sizes, where $100^2$ for example means that we
aligned two sequences, each of
100 character length.
The clolumn titled ``One node'' contains time on a single machine.
The column titled `` 4 nodes with com. time'' reports running times on
four cluster nodes including
communication time. 
The column titled ``4 nodes without com. time'' reports running times on four
cluster nodes 
excluding communication time. 
The column titled ``com. time'' reports communication times only on
four cluster nodes. 
The columns titled ``Win'' and  ``Linux'' contains running times for Windows
and Linux, respectively.
\label{Tab:3}
}
\end{table}

\section*{Acknowledgement}
We would like to thank Nohal Radi and Ahmed Al-Jeshi for their support
in establishing a mutual lab and in “putting in action” the
interoperability “capabilities” between the different systems and
entities involved. We thank also Tamer Shaalan for administrating the
cluster in CMIC and Ayman Kaheel for his support from CMIC. And
special thanks to the Microsoft HPC development team for their useful
guidance and feedback. This project was funded by a grant from
Microsoft.

\bibliographystyle{plain}

\newpage
\section*{Appendix I}
\subsection*{Hardware specs of the cluster at Cairo Microsoft Innovation Center}

We used 
NEXXUS Personal Cluster, model number NEXXUS4080AL, with total memory
of 64 GB and total storage of 1750 GB.
The details are as follows:

\begin{table}[hpb]
\begin{center}
\begin{small}
\begin{tabular}{|c|c|c|}
\hline
\textbf{Item Type}& \textbf{Item Description}&\textbf{Qty}\\
\hline
\hline
{Chassis} & {NEXXUS 4080AL Desk-side Chassis} & {1}\\
\hline
{Compute Shelve} & {Compute Shelve AL/EN (Including Power Supply)} & {4} \\
\hline
{Motherboard} & {Intel S5000AL Motherboard - 1333MHz FSB} & {4}\\
\hline
{Processor} & { Xeon Quad-Core E5410 2.33 12M 1333 MHz - 80W} & {8} \\
\hline
{Memory Type 1 } & {2GB FBDIMMS DDR2 667 ECC/Reg.} & {32}\\
\hline
{Hard Drive  } & {HDD Raid 500GB, 7200rpm, 16MB, SATA II NCQ} & {2}\\
\hline
{Hard Drive  } & {HDD Raid 250GB, 7200rpm, 16MB, SATA II NCQ} & {3}\\
\hline
{Heat Sink  } & {High Performance Passive Heat Sink} & {8}\\
\hline
{Riser Card } & {PCI/Express Riser Card} & {4}\\
\hline
{KVM/USB  } & {Integrated KVM/USB Switch} & {1}\\
\hline
{Gigabit Switch } & {Integrated 16 Port GigE Switch} & {1}\\
\hline
{Power Cord  } & {Power Cord USA/JPN 15Amp ROHS Compliant} & {1}\\
\hline
\end{tabular}
\end{small}
\end{center}
\end{table}

\end{document}